\documentclass[a4paper,preprint,prd,amsmath,floats,tabularx,showpacs,aps,unsortedaddress,superscriptaddress,
nofootinbib]{revtex4}
\usepackage{graphicx}
\usepackage{tabularx}

\begin{document}

\title{Room temperature GW bar detector with opto-mechanical readout}

\author{L.~Conti}
\email{conti@lnl.infn.it}
\address{INFN, Sezione di Padova and Dipartimento di Fisica, Universit\`a di Padova \\ Via Marzolo 8, I-35131 Padova, Italy}

\author{M.~De~Rosa}
\address{INFN, Sezione di Firenze and Dipartimento di Fisica, Universit\`a di Firenze, and LENS\\ Via Sansone 1, I-50019 Sesto Fiorentino (Firenze), Italy}

\author{F.~Marin}
\address{INFN, Sezione di Firenze and Dipartimento di Fisica, Universit\`a di Firenze, and LENS\\ Via Sansone 1, I-50019 Sesto Fiorentino (Firenze), Italy}

\author{L.~Taffarello}
\address{INFN, Sezione di Padova, Via Marzolo 8, I-35131 Padova, Italy}

\author{M.~Cerdonio}
\address{INFN, Sezione di Padova and Dipartimento di Fisica, Universit\`a di Padova \\ Via Marzolo 8, I-35131 Padova, Italy}

\date{\today}

\begin{abstract}
We present the full implementation of a room-temperature gravitational wave bar detector equipped with an opto-mechanical readout. The mechanical vibrations are read by a Fabry--Perot interferometer whose length changes are compared with a stable reference optical cavity by means of a resonant laser. The detector performance is completely characterized in terms of spectral sensitivity and statistical properties of the fluctuations in the system output signal. The new kind of readout technique allows for wide-band detection sensitivity and we can accurately test the model of the coupled oscillators for thermal noise. Our results are very promising in view of cryogenic operation and represent an important step towards significant improvements in the performance of massive gravitational wave detectors.
\bigskip
\end{abstract}

\pacs{07.60.L, 04.80.N, 95.55.Ym, 07.07.M}
\maketitle

\section{Introduction}

The direct observation of gravitational waves (GWs) is one of the most challenging tasks for experimental physics. The effort devoted to this goal started in the 1960's, based on theoretical predictions of the expected signal, which are now considered as very optimistic~\cite{moss71}. 
The experimental strategy for detecting GW signals was based initially on massive acoustic detectors~\cite{barre} and nowadays exploits also long baseline interferometers~\cite{interferometri}. In particular, the former are the most sensitive GW detectors presently in activity~\cite{igec} and they offer interesting possibilities for future advanced versions~\cite{sfera,bisfera}.

Cryogenic bar acoustic detectors are currently equipped with capacitive or inductive transducers followed by SQUID amplifiers or by a microwave resonant cavity. The sensitivity is presently limited by the amplification stage that operates $\sim10^4$ times above the standard quantum limit~\cite{heffner}, giving a bandwidth of few Hz around the two mechanical vibration modes of the coupled oscillators system. 

The possibility of using optical techniques for the readout of bar vibrations was early considered by Drever~\cite{drever77}. Kulagin {\it et al.}~\cite{kulagin86} theoretically studied the possibilities of an optical readout system for a Weber bar with resonant mechanical transformer. This idea was developed by Richard~\cite{richard88}, who designed and theoretically investigated in detail such a system~\cite{richardPRD92}. Richard and coworkers also tested at room temperature an opto-mechanical transducer made by a Fabry--Perot cavity installed on a double oscillator, observing a rms displacement noise consistent with the calculated thermal fluctuations~\cite{richard95}.

In spite of this effort, GW bar detectors have never been equipped with optical readout and the real performance of such a system has never been experimentally verified. On the other hand, the technology has very advanced during the last years in the fields of laser stabilization and fabrication of optical components. As a consequence, the expected characteristics of an optical readout system are even more promising and could lead to a major breakthrough of GW massive detectors.

In the framework of the AURIGA collaboration~\cite{auriga} we are developing a complete optical readout system for ultra-cryogenic bar detectors~\cite{conti98}. The basic idea is to form a high-finesse Fabry--Perot cavity between the bar and a resonant mechanical transducer and then to compare the length of this optical resonator, possibly carrying a GW signal, with that of a stable reference cavity by means of a resonant laser. 

In this work we present the first room temperature GW bar detector operating with an optical readout system. The detector performance is completely characterized in terms of spectral sensitivity and statistical properties of the fluctuations in the system output signal.
Our apparatus represents a wide-band (several tens of Hz), GW acoustic detector which is limited by thermal noise at least in the frequency range of highest sensitivity. Thanks to this property, we can accurately study the output spectrum of the thermal noise and we show that the description of the bar and transducer coupled oscillators cannot be given in terms of de-coupled normal modes, as usually assumed.

The outline of this paper is the following. In Section \ref{sec:readout} we describe the experimental apparatus, and in particular the readout system. In Sec.~\ref{sec:mech} we present the mechanical characteristics and the displacement noise of the detector, as deduced from the output signal. The effect of thermal noise in the output spectrum is analyzed in Sec.~\ref{sec:therm}. Then we investigate the statistical properties of the output fluctuations, performing the accurate analysis necessary to characterize a GW detector (Sec.~\ref{sec:noise}). 
Finally in Sec.~\ref{sec:perform} we estimate the sensitivity of the bar as GW detector. 

\section{The readout system}
\label{sec:readout} 

\begin{figure}
\centering
\includegraphics[width=0.8\columnwidth,clip=]{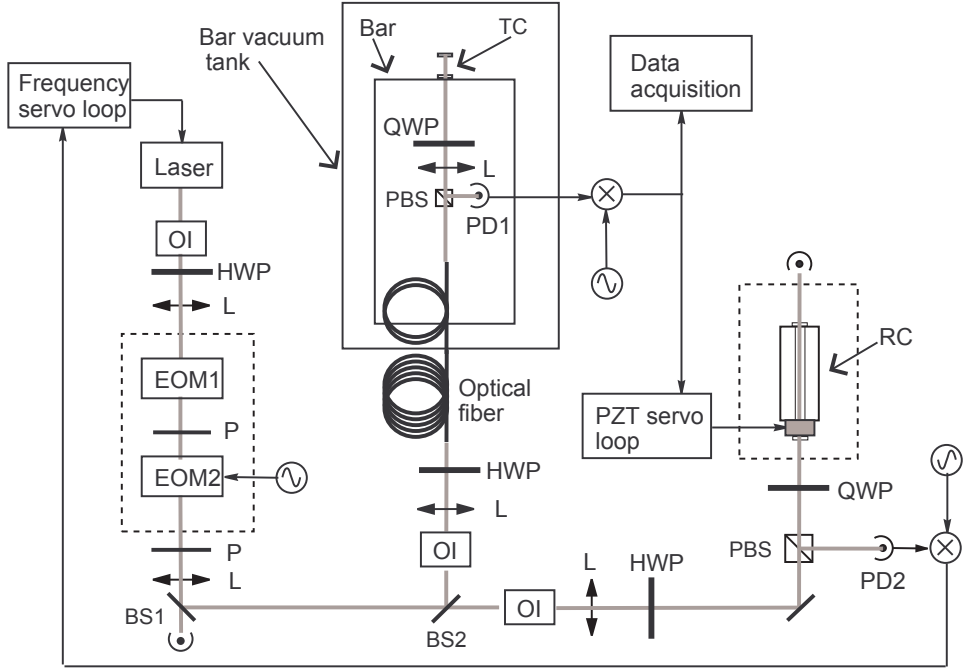}
\caption{Experimental setup: the drawing is not to scale. OI: optical isolator; HWP: half-wave plate: QWP: quarter-wave plate; L: lens; EOM$\#$: electro-optic modulator; P: polarizer; BS$\#$: beam-splitter; PBS: polarizing beam-splitter; PD$\#$: photodiode; RC: reference cavity; TC: transducer cavity. Dashed lines mark out active thermal stabilization.}
\label{fig:schema}
\end{figure}

A schematic drawing of the room-temperature GW detector equipped with the opto-mechanical readout is shown in Fig.~\ref{fig:schema}. The mechanical vibrations of the bar are amplified by a coupled mechanical oscillator and transformed into length changes of an optical cavity, hereafter called transducer cavity (TC). The readout system is composed of a laser source, frequency stabilized to a reference cavity (RC), a set of optical fibers and components to convey the radiation to TC, the opto-electronics for signal detection and elaboration.   

The bar is a 3~m long, 2300~kg mass cylinder made of Al5056. Its first longitudinal vibration mode, useful for GW detection, resonates at 875~Hz, when no load is applied.
The measured resonance frequency varies as $-0.38$~Hz/kg with the amount of non-resonant mass applied to the bar end faces, giving a calculated frequency of $\nu_{\rm b}=866$~Hz when the bar is operated with the full readout system.
The mechanical quality factor of the resonance is $1.8\times 10^5$, measured from the decay time of the excited vibration. The bar is kept in a vacuum chamber, placed a few meters apart from the optical table, and it is isolated from floor mechanical noise by a cascade of passive filters which achieve an overall vertical isolation of about $-$140~dB at the bar frequency, as sensed at the bar middle section. During the work here reported the vacuum system, composed by a roots pump backed by a rotary pump, was operated just for about 1 hour per day.

The output beam of a commercial Nd:YAG laser source, emitting 50~mW at 1.064~$\mu$m, passes through an optical isolator and two electro-optic modulators (EOMs) enclosed in a thermally stabilized box. The first EOM is used with a polarizer in an amplitude stabilization loop. The purpose of this noise-eater, described in detail in Ref.~\cite{AP_ottico}, is reducing the effect of the back-action on the transducer, but it is not relevant for the work here reported. The second EOM is a resonant modulator working at 13.3~MHz which accomplishes phase modulation with a depth of about 1~rad. A first beam-splitter (BS1) transmits 20$\%$ of the radiation for the noise eater, while the reflected beam is directed towards a second 70$\%$ transmission beam-splitter (BS2). The reflected beam, after an optical isolator, is coupled to a single-mode polarization-maintaining fiber and arrives to a $135 \times 340$~mm$^2$ Al plate anchored to the bar middle section. 

The optical fiber is formed by joining 4 patchcords with FC/PC connectors, for a total length of 13~m, and includes a homemade vacuum feedthrough. The two fiber ends have pigtailed collimators, with anti-reflection coating. The overall power transmission of the fiber assembly is about 50$\%$.

The 1.5~mW collimated beam transmitted by the fiber passes through an optical circulator, formed by a polarizing beam-splitter and a quarter-wave plate, and a telescope to properly couple the radiation to the TEM$_{00}$ mode of TC. Four tilting mirrors send the beam towards the transducer cavity on the bar end face. The beam reflected by TC, after the circulator, is detected by a photodiode (PD1).
The TC is a 6 mm long Fabry--Perot cavity, with a finesse of 28000, formed by an input concave mirror (radius of curvature 1~m, diameter $0.5^{\prime\prime}$) glued to a support fixed to the bar, and a flat back mirror (diameter $0.5^{\prime\prime}$) glued to the oscillating mass of the mechanical transducer. This resonant transducer is machined from a single piece of Al5056 and it is composed of a thin circular plate loaded by a central 1.25~kg inert mass. The resonant frequency of the first drum mode is about 882~Hz, according to the measurements described in Sec.~\ref{sec:mech}. This resonator was designed for a previous version of the readout system and it is described in Ref. \cite{conti98}. 

The beam transmitted by BS2, after an optical isolator and mode-matching lenses, is sent to a 110~mm long Fabry--Perot reference cavity that has a finesse of 44000. RC is formed by an Invar spacer with a couple of mirrors similar to the ones of the TC. The input flat mirror is glued on a piezoelectric actuator (PZT), which allows the tuning of the cavity length. The light power impinging on the cavity is about 4.5~mW. The cavity is kept in a vacuum chamber whose temperature is actively stabilized at about 34$^\circ$C within 0.1$^\circ$C. The beam reflected by this cavity is detected by a second photodiode (PD2) after an optical circulator. 

The power level impinging on PD1 shows large variations: during the 42 hour period of continuous data acquisition, it varied by up to 80$\%$, on a time-scale of typically a few hours. On the same period the power impinging on PD2 varied by less than 10$\%$, with a longer timescale. The large variations sensed by PD1 are due to polarization fluctuations generated by drifts of the room temperature and originating from a non-perfect matching between the polarization axes of the fiber patchcords. They are turned into changes of the power impinging on TC by the optical circulator at the fiber output. 

The ac component of the signals coming from the two photodiodes is demodulated at 13.3~MHz and filtered, according to the Pound--Drever scheme~\cite{Drever}.
The resulting signals are used as discriminator for frequency locking and analysis. The laser frequency is locked to a resonance peak of RC with a servo loop which has a unit gain frequency of 30~kHz. The loop gain is maximized in the frequency range between 600~Hz and 900~Hz, where it reaches 120~dB, and allows to achieve an in-loop frequency noise level below the shot noise limit. A detailed description of the loop electronics and performance can be found in Ref.~\cite{Conti-phd}. 
The resonant peak of RC is then superimposed to a resonance of TC by operating on the PZT actuator. The Pound--Drever signal from TC is used in a servo loop which drives the PZT through a low noise high-voltage amplifier. This low frequency servo loop has unity gain at about 1~Hz and allows the reference cavity to follow the resonance of TC in its thermal drifts. Due to the low loop bandwidth, the two cavities can be considered as free and independent in the frequency range of interest. 

The same error signal from TC is acquired and analyzed to extract information concerning the motion of the bar detector.

\section{Experimental results} 
\label{sec:exp}

\subsection{Mechanical system and noise spectrum}
\label{sec:mech}

\begin{figure}
\centering
\includegraphics*[bbllx=30bp,bblly=20bp,bburx=725bp,bbury=510bp,width=0.8\columnwidth]{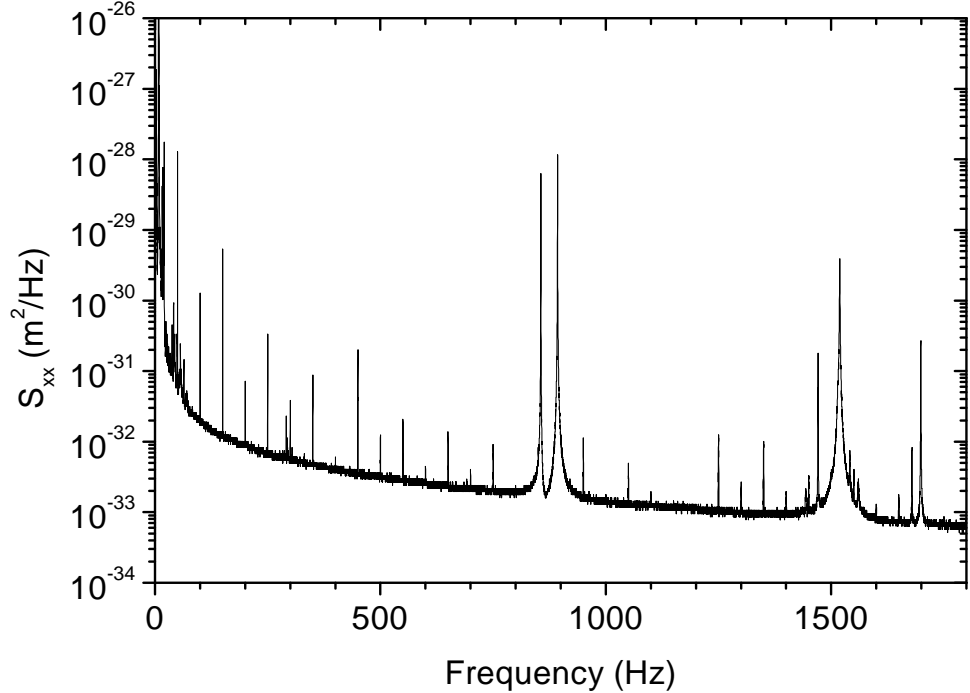}
\caption{Power spectral density $S_{xx}$ of the displacement noise. The two peaks at 856~Hz and 892~Hz corresponds to the two mechanical modes of coupled oscillators system formed by the bar with its first longitudinal mode and the transducer. Some other mechanical resonances are visible too. The sharp peaks at multiples of 50~Hz are due to the power line.}
\label{fig:noise}
\end{figure}

The conversion of the detector output signal from voltage into length change of the transducer cavity is obtained using the slope of the corresponding error signal, measured with an accuracy of about 20$\%$, and the known cavity length. The power spectral density $S_{xx}$ of the displacement noise is shown in Fig.~\ref{fig:noise}, which corresponds to the average of one hour data. The peaks at 856~Hz and 892~Hz correspond to the frequencies $\nu_\pm$ of the two mechanical modes of the coupled oscillators system formed by the bar and the resonant transducer.
For the coupled system the following relation holds: $\nu_+ \nu_- = \nu_{\rm b} \nu_{\rm t}$, where $\nu_{\rm t}$ is the resonance frequency of the transducer.
According to the value of $\nu_{\rm b}=866$~Hz calculated for the loaded bar we infer that the transducer resonance is at 882~Hz and thus detuned by +16~Hz with respect to the bar. 
A better coupling between the two oscillators is possible by adjusting the thickness of the transducer circular plate.

We determined the mechanical quality factor $Q$ of the `+' and `$-$' modes by measuring the decay time of a resonant sinusoidal excitation applied to the bar by means of a piezoelectric actuator situated at the bar end face opposite the transducer. We got $Q_-=16600$ and $Q_+=8700$. The factor of 2 of difference in the $Q$s is explained by considering that, due to the frequency detuning between the two resonators, the minus mode is more influenced by the (high $Q$) bar while the plus mode by the (low $Q$) transducer.

As it can be seen in Fig.~\ref{fig:noise}, the peak spectral power in the modes exceeds by about 45~dB the background noise. 
The wideband output noise of the optical readout system comes from the residual frequency fluctuations of the laser stabilized to RC and from the noise measured when the laser is far from the TC resonance. 
The laser frequency noise has been measured with respect to a stable Zerodur cavity, with an apparatus described in Ref.~\cite{JOSA}, and its effect gives a sensitivity limit as low as $2 \times 10^{-2}$~Hz/$\sqrt{\rm Hz}$ around 1~kHz. The far-from-resonance fluctuations are due to electronic noise, laser amplitude noise (including shot noise), and interference fringes. The overall effect gives a sensitivity limit of about 0.2~Hz/$\sqrt{\rm Hz}$. The observed background, visible in Fig.~\ref{fig:noise}, is about 20~dB higher than this limit and exhibits a decreasing behavior versus frequency~\cite{Amaldi}.

We fitted the noise power spectral density of one hour output data with the function $A/\nu^{\epsilon}$ between 75~Hz and 1775~Hz, neglecting only the interval around the `+' and `$-$' modes and the resonance at 1519~Hz.  We obtained $\epsilon=1.16$ and $A=4.2 \times 10^{-30}$~m$^2$Hz$^{\epsilon}$/Hz. The origin of this noise is unknown and is presently under investigation. We remark that a $1/f$ frequency dependence in the noise spectrum is expected for the thermal noise of a mechanical oscillator with internal friction modeled as a constant imaginary part of the spring constant, in the frequency region below the oscillator resonance.
 
\subsection{Thermal noise of the coupled oscillators system}
\label{sec:therm}

A bar detector equipped with a resonant transducer is widely modeled as a system of two coupled harmonic oscillators and the output is often analyzed in terms of normal mode expansion. In particular, such assumption is used for studying the thermal noise of the system~\cite{Saulson}. On the other hand, it has been suggested~\cite{Majorana} and experimentally verified~\cite{jap} that this description may fail if inhomogeneously distributed losses occur. For our system the condition for the validity of the normal mode expansion can be written, according to Ref.~\cite{Majorana}, as 
\begin{equation}
\nu_{\rm b} Q_{\rm b} = \nu_{\rm t} Q_{\rm t} \, .
\label{condizione}
\end{equation}
In our case, the large difference between the quality factors of the modes `+' and `$-$' can be brought back to a large difference in the effective $Q$s of the original oscillators $Q_{\rm b}$ and $Q_{\rm t}$. Therefore, since the frequencies $\nu_{\rm b}$ and $\nu_{\rm t}$ are very similar, the condition of Eq.~(\ref{condizione}) is not satisfied. 

We call $x_{\rm b}$, $m_{\rm b}$ and $x_{\rm t}$, $m_{\rm t}$ the coordinate and effective mass of the bar and transducer oscillator respectively, and $f_{\rm b}$ and $f_{\rm t}$ the corresponding total driving force. Assuming that only viscous damping is present,\footnote{We also considered the case of structural losses as dissipative mechanism, but no difference was found within the reported errors.}
the dynamics of the system is described by the equations of motion
\begin{equation}
\begin{cases}
\displaystyle
\ddot{x_{\rm b}} + \frac{\omega_{\rm b}}{Q_{\rm b}} \, \dot{x_{\rm b}} + \mu \, \frac{\omega_{\rm t}}{Q_{\rm t}} \, (\dot{x_{\rm b}}-\dot{x_{\rm t}})  + \omega_{\rm b}^2 \, x_{\rm b} + \mu \, \omega_{\rm t}^2 \, (x_{\rm b} - x_{\rm t})=  \frac{f_{\rm b} - f_{\rm t}}{m_{\rm b}} \nonumber \\
\\
\displaystyle
\ddot{x_{\rm t}} + \frac{\omega_{\rm t}}{Q_{\rm t}} \, (\dot{x_{\rm t}}-\dot{x_{\rm b}}) + \omega_{\rm t}^2 \, (x_{\rm t}-x_{\rm b}) = \frac{f_{\rm t}}{m_{\rm t}} \; \; , \nonumber 
\end{cases} 
\end{equation}
where $\mu = m_{\rm t}/m_{\rm b}$ and $\omega_{\rm b,t} = 2\pi \nu_{\rm b,t}$. In the frequency domain the system can be written as
\begin{equation}
{\bf D} (\omega) \begin{pmatrix} X_{\rm b} \\ X_{\rm t} \end{pmatrix} = \begin{pmatrix} (F_{\rm b}-F_{\rm t})/m_{\rm b} \\ F_{\rm t}/m_{\rm t} \end{pmatrix} \, ,
\end{equation}
with 
\[{\mathbf D} (\omega)= \begin{pmatrix} -\omega^2 + \omega_{\rm b}^2 + \mu \, \omega_{\rm t}^2 + i \, \omega \left( \displaystyle\frac{\omega_{\rm b}}{Q_{\rm b}}+ \mu \, \frac{\omega_{\rm t}}{Q_{\rm t}}\right) & - \mu \, \omega_{\rm t}^2 - i \, \mu \displaystyle\frac{\omega \, \omega_{\rm t}}{Q_{\rm t}} \\
 & \\
- \omega_{\rm t}^2 - i \,\displaystyle \frac{\omega \, \omega_{\rm t}}{Q_{\rm t}} & -\omega^2 + \omega_{\rm t}^2 + i \, \displaystyle\frac{\omega \, \omega_{\rm t}}{Q_{\rm t}}
\end{pmatrix} \, ,
\]
where capital letters indicate Fourier transforms and $i$ is the imaginary unit.

To simplify our analysis, in the following we consider the transducer cavity length changes as determined exclusively by a motion of the transducer. This is justified as the amplitude of a bar displacement is amplified at the transducer by a factor equal to $1/\sqrt{\mu}$, i.e., by a factor of about 30. We are thus interested to the noise power spectral density $S_{x_{\rm t}x_{\rm t}}$ of $x_{\rm t}$ which, if only stochastic thermal forces are present, can be written according to the Fluctuation-Dissipation Theorem~\cite{Majorana} as
\begin{equation}
S_{x_{\rm t}x_{\rm t}}^T(\omega)= \frac{2 k_{\rm B} T}{\omega^2} Re \{ (i \omega {\bf D})^{-1} \, _{22} \} \, ,
\label{eq:FDTfit}
\end{equation}
where $k_{\rm B}$ is the Boltzmann constant and $T$ is the thermodynamic temperature. In order to account for the observed $\sim 1/f$ background, we add a phenomenological wide-band term:
\begin{equation}
S_{x_{\rm t}x_{\rm t}}(\omega)= S_{x_{\rm t}x_{\rm t}}^T(\omega) + \frac{S_{\rm WB}}{\omega^\epsilon} \, . 
\label{eq:FDTfit2}
\end{equation}

\begin{figure}
\includegraphics*[bbllx=30bp,bblly=30bp,bburx=725bp,bbury=510bp,width=0.8\columnwidth]{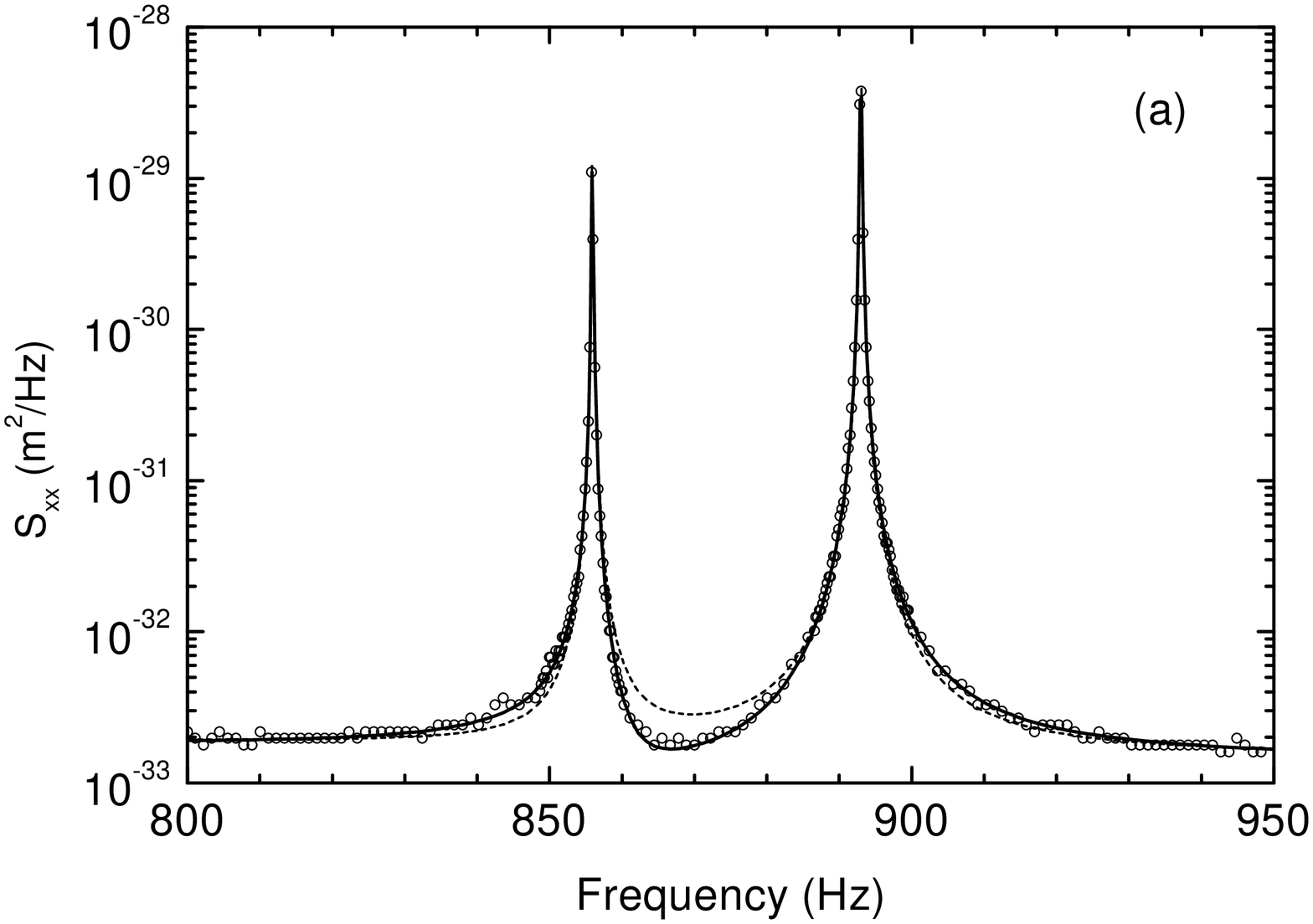}
\includegraphics*[bbllx=30bp,bblly=30bp,bburx=725bp,bbury=510bp,width=0.8\columnwidth]{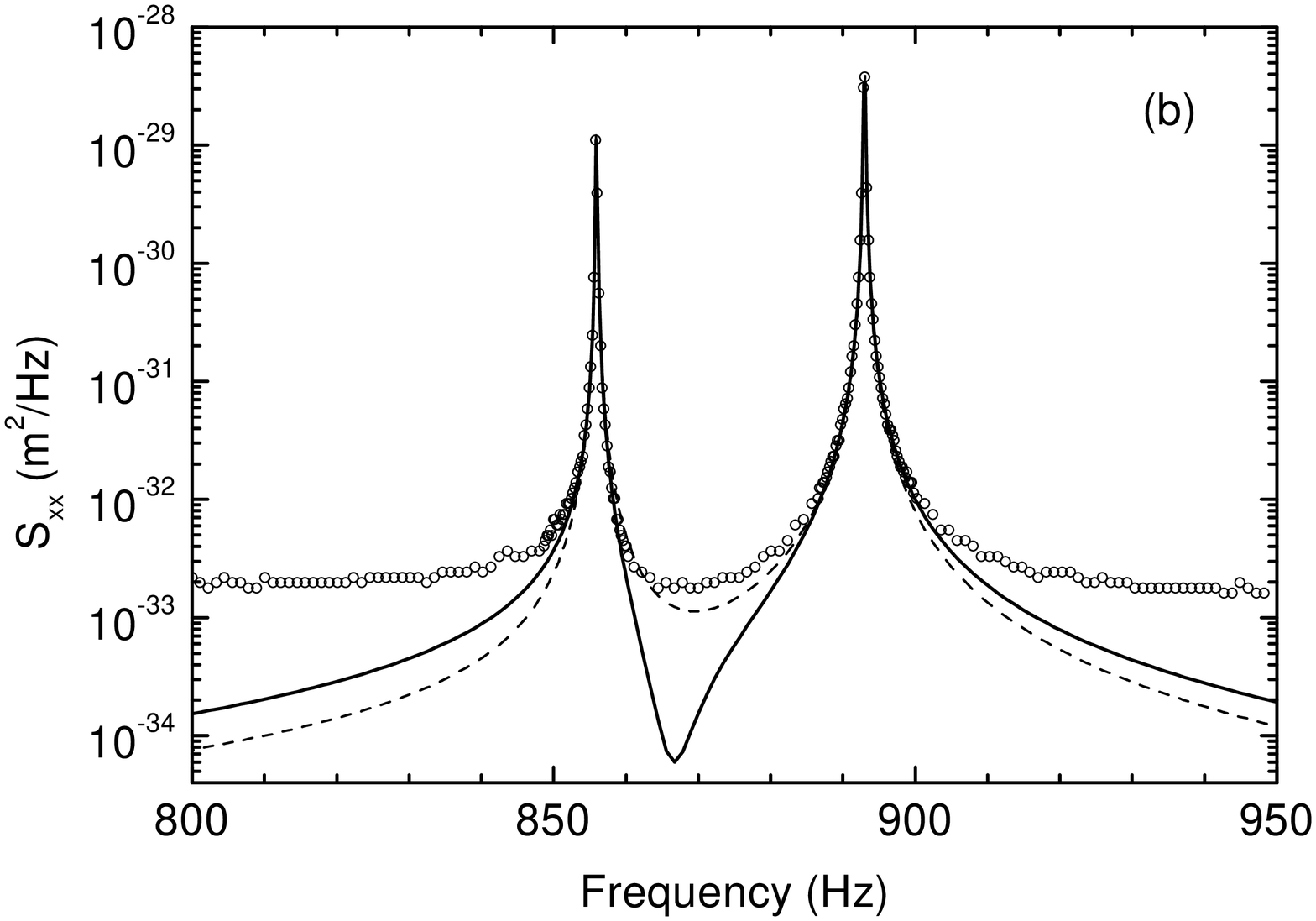}
\caption{a) Power spectral density of the displacement noise. Circles: experimental data; solid line: fit according to the two-oscillators model of Eqs.~(\ref{eq:FDTfit}) and (\ref{eq:FDTfit2}); dashed line: fit according to the normal mode expansion. b) The same as in a), where the fitted curves are plotted without the wide-band $\sim 1/f$ contribution.}
\label{fig:fit}
\end{figure}

We fitted the output spectrum using the expression of Eq.~(\ref{eq:FDTfit2}), in the frequency range between 780~Hz and 950~Hz. The fit allows to infer the ratios $Q_{\rm b}/T$, $Q_{\rm t}/T$ and the resonant frequencies $\nu_{\rm b}$, $\nu_{\rm t}$ of the uncoupled oscillators, the effective mass $m_{\rm t}$ of the transducer oscillator and the magnitude $S_{\rm WB}$ of the background noise. The effective mass $m_{\rm b}=1180$~kg of the loaded bar resonator and the temperature $T$ were kept constant during the fitting. We assumed that both oscillators are at the same thermodynamic temperature $T$ of 296~K, as measured by a probe placed on the bar. This assumption seems reasonable at least for the bar as the inferred value for $Q_{\rm b}$ agrees very well with the one measured independently for the unloaded bar ($1.8\times 10^5$). The result of the fitting procedure is shown in Fig.~\ref{fig:fit} and the parameters are summarized in Tab.~\ref{Table_1}.
The two frequencies $\nu_{\rm b}$ and $\nu_{\rm t}$ agree well with the values estimated in Secs.~\ref{sec:readout} and \ref{sec:mech}. Also the scale factor $S_{\rm WB}$ of the background noise agrees with the value $(2\pi)^\epsilon A=3.5 \times 10^{-29}$~m$^2$Hz$^{\epsilon}$/Hz obtained in Sec.~\ref{sec:mech}. As far as the transducer mass is concerned we notice that its effective mass deduced from the fit is greater than the mere 1.25~kg central mass.  

We also attempted to fit the same data with the prediction of the normal mode expansion, as shown in Fig.~\ref{fig:fit}. It is evident that, while the experimental spectrum is in excellent agreement with the model of Eqs.~(\ref{eq:FDTfit}) and (\ref{eq:FDTfit2}), the normal mode expansion fails to describe our system: the presence of  inhomogeneously distributed losses causes the random fluctuations of the two initial oscillators to be correlated. The normal mode expansion overestimates the noise in between the two modes because it does not take into account such a correlation.

\begin{table}[b]
\caption{Results of the fit of the experimental spectrum with Eqs.~(\ref{eq:FDTfit}) and (\ref{eq:FDTfit2}), assuming $T$=296~K for both resonators. The quoted errors (two standard deviations) refer to the last significant digit.}\label{Table_1}
\centering
\begin{tabularx}{0.7\columnwidth}{XXX}
\hline\hline 
Parameter & Fitted value & units \\
\hline
$\nu_{\rm b}$ & 866.31 (3)&Hz \\
$\nu_{\rm t}$ & 882.30 (3)&Hz \\ 
$Q_{\rm b}$ & $1.8 \times 10^5$ (4) \\
$Q_{\rm t}$ & $6.60 \times 10^3$ (4) \\ 
$S_{\rm WB}$ & $3.5 \times 10^{-29}$ (2) &m$^2$Hz$^{-\epsilon}$/Hz \\
$m_{\rm t}$ & 1.70 (2) &kg \\ 
\hline\hline
\end{tabularx}
\end{table}

\subsection{Statistical analysis of the output fluctuations}
\label{sec:noise}

For a deep understanding of the detection system, it is important to investigate the statistical behavior of the noise before attempting any estimation of its magnitude. In fact, one needs to be sure that the noise under observation follows the laws predicted for the expected noise sources. In particular, the fundamental hypothesis is that the noise is a (quasi-)stationary stochastic process with Gaussian statistics. The system statistics is a crucial issue for GW detectors and it can cause a dramatic decrease of the effective duty cycle. Indeed, it is safe to limit the analysis only to the periods when the experimental noise is well modeled, indicating that the detector is working properly.

The output of PD1 was recorded with an acquisition system identical to the one employed for the ultra-cryogenic GW detector AURIGA: the data are sampled at 4.88~kHz and synchronized to UTC with a GPS clock. We have acquired data between May 31st 2001 and June 20th 2001. The data acquisition was not continuous due to intentional interruptions for diagnostic purposes and to system failures mainly originated by environmental temperature variations~\cite{Amaldi}. Manual relocking procedures require less than 15 minutes and the longest continuous locking period was 42 hours. The overall data recording corresponds to 183 hours. A few additional signals were sampled at 20~Hz for monitoring the dc signals from PD1 and PD2, the temperature of the RC and the correction voltage fed to the PZT of the RC.

The acquired data were elaborated through the same data analysis used for the ultra-cryogenic detector~\cite{analisi_au}. The analysis implements a Wiener-Kolmogorov (WK) filter to search for $\delta$-like signals ({\em triggers}), i.e., for short bursts whose Fourier transform can be considered as constant over the effective bandwidth of the detector. A maximum-hold algorithm is applied to the data and for each trigger we estimate the time of arrival, the amplitude and the $\chi^2$ with respect to the expected shape. The latter discriminates a $\delta$-like mechanical excitation of the bar, i.e., the GW signal, from spurious signals. The analysis also implements adaptive algorithms that update the parameters of the WK filter in order to follow slow drifts of the system. 

\begin{figure}
\centering
\includegraphics[width=0.8\columnwidth,clip=]{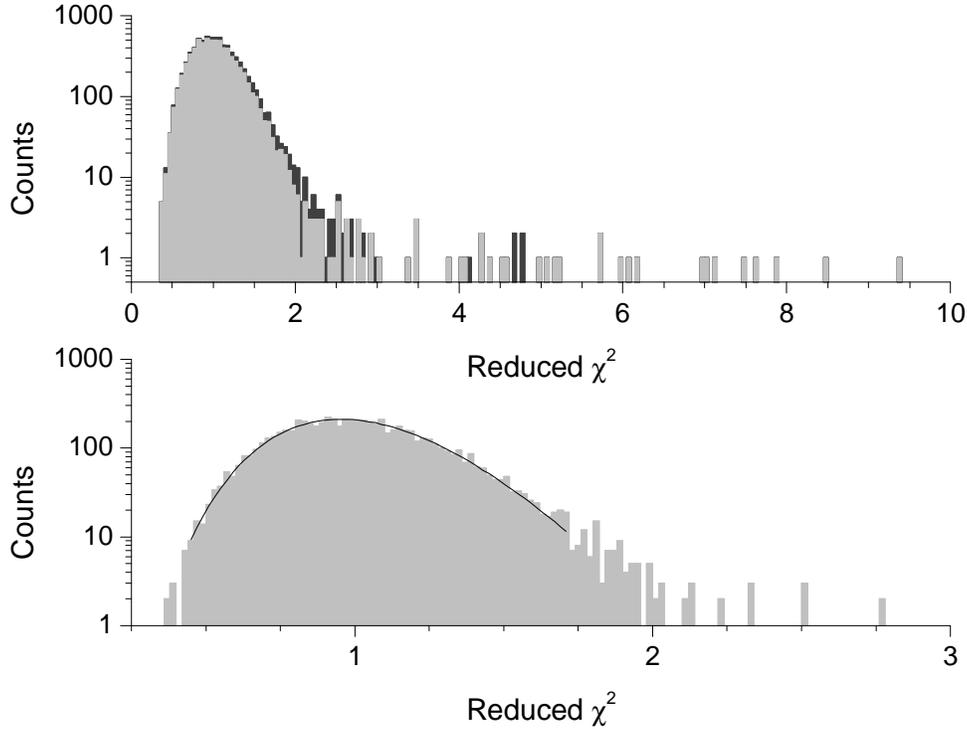}
\caption{Top: histogram of $\chi^2_a$ of all triggers with SNR$>$4, before (dark gray) and after veto (light gray). Bottom: histogram of all triggers after veto and with 4$<$SNR$<$6. The solid line is the reduced $\chi^2$ distribution with 30 degrees of freedom fitted to the vetoed data with 4$<$SNR$<$6. The data refer to June 9th 2001, and the veto is applied when vacuum pumps are on. The binning corresponds to intervals of 0.005 for $\chi^2_a$.}
\label{fig:chi}
\end{figure}

In order to verify the Gaussian behavior of the detection output signal, we study the distribution of the reduced $\chi^2$ (called $\chi^2_a$ in the following) of all triggers found by the data analysis: for a random variable with Gaussian statistics $\chi^2_a$ should follow a well-known distribution~\cite{distrib_chi}. We focus on the 24 hours of data acquired during June 9th. The pumps that maintain the vacuum in the bar tank were switched on for one hour, between h12.30 UTC and h13.30 UTC. We have vetoed the data acquired within this period in order to avoid the effect of the noise introduced by the pumps. We plot in Fig.~\ref{fig:chi} (top) the histogram of $\chi^2_a$ of all triggers with signal-to-noise ratio (SNR) greater than 4~\cite{ortolan}, before and after applying the veto (respectively, dark and light gray histogram). The tails at higher values of $\chi^2_a$ are very efficiently cut away by considering only those triggers with 4$<$SNR$<$6 (bottom histogram). The number of degrees of freedom used to compute $\chi^2_a$ is 30. We can well fit the data having 4$<$SNR$<$6, after veto, with the theoretical distribution of $\chi^2_a$: the reduced $\chi^2$ of the fit is 0.90.

\begin{figure}
\centering
\includegraphics[width=0.8\columnwidth,clip=]{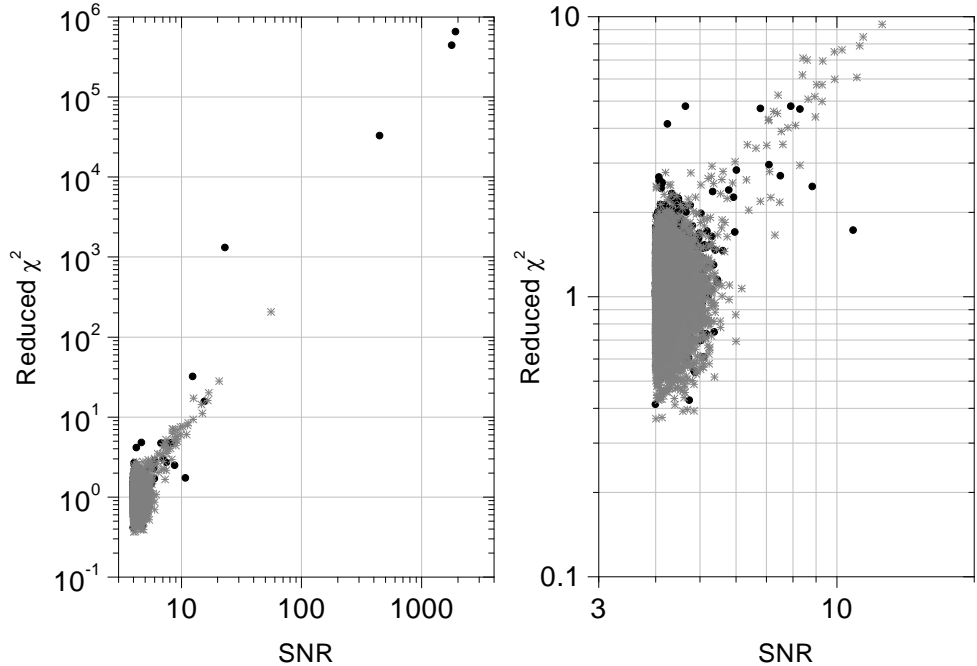}
\caption{Scatter plot of $\chi^2_a$ versus SNR for all triggers after veto and with SNR$>$4 (gray stars). The closed circles correspond to the vetoed triggers. The data refers to June 9th 2001, and the veto is applied when vacuum pumps are on.}
\label{fig:snrchi}
\end{figure}

In Fig.~\ref{fig:snrchi} we plot the $\chi^2_a$ of all triggers with SNR$>$4 versus the SNR. Most of them concentrate in the region of low SNR and $\chi^2_a<2$. A small fraction (about 0.05$\%$) of triggers follows a linear law in the log-log plot of Fig.~\ref{fig:snrchi}. Indeed, it has been shown~\cite{analisi_au} that a quadratic scale law of $\chi^2_a$ versus the SNR is expected for signals which are not matched by the filter. 92$\%$ of triggers with SNR$>6$ surviving the veto can be rejected as they have a $\chi^2_a>2.1$, a threshold that corresponds to a confidence level of $3.9 \times 10^{-4}$ for our 30 degrees of freedom.

\begin{figure}
\includegraphics[width=0.8\columnwidth,clip=]{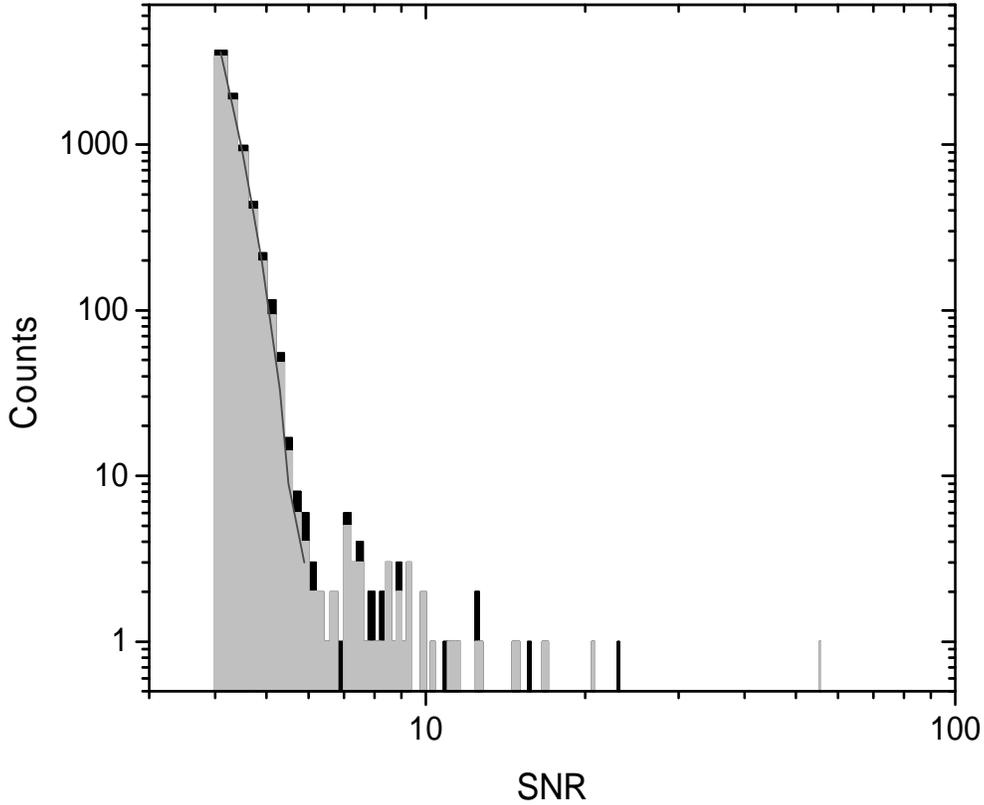}
\caption{Histogram of the SNR of all triggers with SNR$>$4, before (black) and after veto (light gray). The data refers to June 9th 2001, and the veto is applied when vacuum pumps are on. The solid line is the distribution predicted by a numerical simulation of a Gaussian process with the same parameters as the detection system. The binning corresponds to intervals of 0.2 for SNR.}
\label{fig:snr}
\end{figure}

An independent test of the Gaussian character of the system is the distribution of the SNR. In this case the analytical formula for the expected distribution involves the calculation of integrals that are not easily solvable. We therefore compare the experimental distribution of SNR with that obtained by simulating with Monte Carlo methods a Gaussian system having the same parameters as ours (namely, frequencies, bandwidth and ratio between the height of the mechanical mode peaks and the background noise). The simulation output is passed through the same analysis as the real data, so that any deviation of our system from a Gaussian behavior would appear as a difference in the SNR distribution between the simulated and the true system. The results are plotted in Fig.~\ref{fig:snr} and the agreement is excellent, above all considering that 92$\%$ of the triggers at SNR$>6$ are rejected by the $\chi^2$ test. The remaining signal distribution perfectly corresponds to the expected output of a system with Gaussian input noise, without any excess trigger.

\begin{figure}
\includegraphics*[bbllx=50bp,bblly=30bp,bburx=740bp,bbury=500bp,width=0.8\columnwidth,clip=]{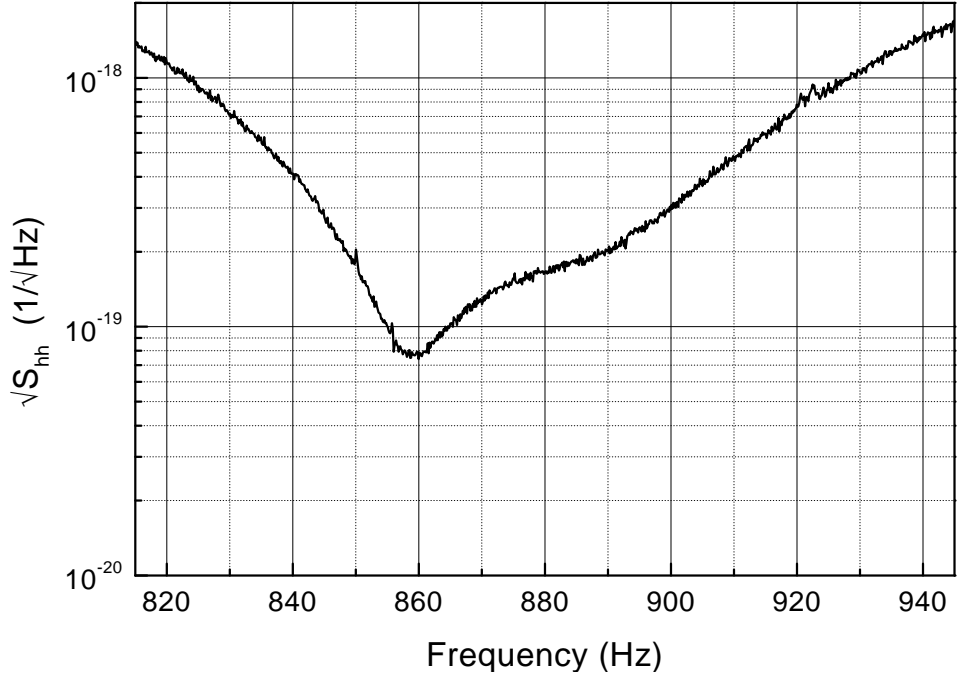}
\caption{Sensitivity of the GW detector, expressed in terms of the equivalent strain noise at the detector input. The data correspond to a one-hour average.}
\label{fig:shh}
\end{figure} 

\subsection{Performance as GW detector}
\label{sec:perform}

The sensitivity of the bar detector is determined from $S_{hh}$, defined as the spectral density of the total noise referred to the detector input and calibrated in terms of GW amplitude. Figure~\ref{fig:shh} shows $\sqrt{S_{hh}}$ in a neighbourhood of the two modes `+' and `$-$'. As expected, the best sensitivity is achieved in correspondence of the modes, with a peak at the mode `$-$', that has a higher $Q$. 

A critical parameter for a GW resonant detector is the detection bandwidth, which is particularly significant for the temporal definition of the candidate events and for the synchronization of several detectors~\cite{timing}. The spectral sensitivity of the operating cryogenic bar detectors is characterized, with only one exception~\cite{explorer}, by narrow quasi-Lorentzian peaks, due to the relatively large amplifier noise. In that case, the detection bandwidth is of the order of few Hz. For our detector, the sensitivity is relevant in the whole frequency interval between the modes, as expected for an optimized resonant transducer. The significant bandwidth exceeds the modes splitting. The output spectrum is asymmetric, but we remark that at 10~dB from the minimum the width is about 50~Hz.

The amplitude of a GW burst that would be detected with unitary SNR is $h_{\min}=3 \times 10^{-17}$. This is equivalent to a standard pulse of $0.8 M_\odot c^2$ converted into GW at the distance of 10~kPc, i.e., at the galactic center. This performance is not of astrophysical interest, as expected for the room temperature operation,  but it is very promising: once the system is upgraded both in the optics and in the detector design to operate at ultra-cryogenic temperatures, the sensitivity is expected to increase up to $h_{\min} \sim 10^{-20}$.

\section{Conclusions}

We have operated a room temperature GW bar detector equipped with an opto-mechanical readout. The sensitivity is limited by thermal noise, due to the high temperature and low mechanical quality factors. Both parameters will improve in the case of cryogenic operation. The statistics of the detector noise is stable and Gaussian as expected. The sensitivity is enough to evidence the failure of the normal mode expansion due to the presence of inhomogeneously distributed losses. The achieved results are a significant step towards the realization of an ultra-cryogenic resonant detector equipped with optomechanical readout. This is predicted to improve the bandwidth of the AURIGA detector by more than one order of magnitude with respect to its present status.

\begin{acknowledgments}
We gratefully acknowledge all the former and current components of the AURIGA group without whom this work would not have been done. We thank especially G.~A.~Prodi and J.~P.~Zendri for helping in the day by day laboratory work, L.~Baggio, A.~Ortolan and G.~Vedovato for providing the data acquisition system and the data analysis tools and adapting them to our detector parameters.

This work was partially funded by the MURST (research program `Transducer systems for cryogenic resonant detectors of gravitational waves').

\end{acknowledgments}


\begin{references}
\bibitem{moss71} G.E. Moss, L.R. Miller, and R.L. Forward, Appl. Opt. {\bf 10}, 2495 (1971) and references therein.

\bibitem{barre}
P.~Astone, Class. Quant. Grav. {\bf 19}, 1227 (2002).

\bibitem{interferometri}
B. Willke {\itshape et al.}, 
Class. Quant. Grav. {\bf 19}, 1377 (2002);
D. Sigg {\itshape et al.}, 
{\it ibid.} {\bf 19}, 1429 (2002);
Masaki Ando {\itshape et al.},
{\it ibid.} {\bf 19}, 1409 (2002);
F. Acernese {\itshape et al.},
{\it ibid.} {\bf 19}, 1421 (2002).

\bibitem{igec} Z.A.\ Allen {\itshape et al.},  Phys.\ Rev.\ Lett. {\bf 85}, 5046 (2000). \url{http://igec.lnl.infn.it/igec}.

\bibitem{sfera} 
E.~Coccia {\itshape et al.}, Phys. Rev. D {\bf 57}, 2051 (1998).

\bibitem{bisfera} 
M.~Cerdonio {\itshape et al.}, Phys. Rev. Lett. {\bf 87}, 031101 (2001).

\bibitem{heffner} H.\ Heffner, Proc. IRE {\bf 45}, 1604 (1962).

\bibitem{drever77} 
R.~W.~P.~Drever {\itshape et al.}, in {\it Proceedings of the International Meeting of Experimental Gravitation}, edited by B.~Bertotti (Atti Convegni Lincei, vol.~34, 1977) p.~365.

\bibitem{kulagin86} V.V. Kulagin, A.G. Polnarev, and V.N. Rudenko,  Sov. Phys. JEPT {\bf 64}, 915 (1986).

\bibitem{richard88} J.-P.\ Richard, J. Appl. Phys. {\bf 64}, 2202 (1988).

\bibitem{richardPRD92} J.-P.\ Richard,  Phys.\ Rev.\ D {\bf 46}, 2309 (1992). 

\bibitem{richard95} Yi Pang and J.-P.\ Richard, Appl. Opt. {\bf 34}, 4982 (1995).

\bibitem{auriga}
G.A. Prodi {\em et al.}, in {\it Gravitational Waves, Proceedings of the second Edoardo Amaldi Conference}, edited by E.~Coccia {\em et al.} (World Scientific, 1998), p.~148. J.~P.~Zendri {\em et al.}, Class. Quant. Grav. {\bf 19}, 1925 (2002).
For more information, see also \url{http://www.auriga.lnl.infn.it}. 

\bibitem{conti98}  L.\ Conti {\it et al.}, Rev. Sci. Instrum. {\bf 69}, 554 (1998)

\bibitem{AP_ottico} L.\ Conti, M.\ De\ Rosa and F.\ Marin, Appl.\ Opt. {\bf 39}, 5732 (2000).

\bibitem{Drever}
R.~W.~P.~Drever {\em et al.}, Appl. Phys. B {\bf 31}, 97 (1983).

\bibitem{Conti-phd}
L. Conti, PhD Thesis, University of Trento, (1999). Downloadable at the AURIGA web site: \url{http://www.auriga.lnl.infn.it/publications/publications.html}.

\bibitem{JOSA}
L. Conti, M. De Rosa and F. Marin, submitted to Journal of the Optical Society of America B.

\bibitem{Amaldi} 
M.~De~Rosa {\em et al.}, Class. Quant. Grav. {\bf 19}, 1919 (2002).

\bibitem{Saulson} 
P.~R.\ Saulson, Phys.\ Rev.\ D {\bf 42}, 2437 (1990).

\bibitem{Majorana} E.\ Majorana and Y.\ Ogawa, Phys.\ Lett.\ A {\bf 233}, 162 (1997).

\bibitem{jap} K.\ Yamamoto, S.\ Otsuka, M.\ Ando, K.\ Kawabe, K.\ Tsubono,  Phys.\ Lett.\ A {\bf 280}, 289 (2001).

\bibitem{analisi_au} L.~Baggio {et al.}, Phys.\ Rev. D {\bf 61}, 102001 (2000) and references therein.

\bibitem{distrib_chi} See for instance A.\ Papoulis, {\it Probability, Random Variables, and Stochastic Processes}, 3rd ed., McGraw-Hill, Singapore (1991) p.79.

\bibitem{ortolan} A.\ Ortolan {\it et al.}, Class. Quant. Grav. {\bf 19}, 1457 (2002).

\bibitem{timing} V.\ Crivelli\ Visconti {\it et al.}, Phys.\ Rev. D {\bf 57} 2045 (1998).

\bibitem{explorer} P.\ Astone {\it et al.}, Class. Quant. Grav. {\bf 19}, 1905 (2002).


\end{references}
\end{document}